# Towards Enabling Critical mMTC: A Review of URLLC within mMTC

Shiva Raj Pokhrel, Jie Ding, Jihong Park, Ok-Sun Park and Jinho Choi

*Abstract*—Massive machine-type communication (mMTC) and ultra-reliable and low-latency communication (URLLC) are two key service types in the fifth-generation (5G) communication systems, pursuing scalability and reliability with low-latency, respectively. These two extreme services are envisaged to agglom- erate together into critical mMTC shortly with emerging use cases (e.g., wide-area disaster monitoring, wireless factory automation), creating new challenges to designing wireless systems beyond 5G. While conventional network slicing is effective in supporting a simple mixture of mMTC and URLLC, it is difficult to simultaneously guarantee the reliability, latency, and scalability requirements of critical mMTC (e.g., < 4ms latency, $10_6$ devices/km$^2$ for factory automation) with limited radio resources. Furthermore, recently proposed solutions to scalable URLLC (e.g., machine learning aided URLLC for driverless vehicles) are ill-suited to critical mMTC whose machine type users have extremely limited energy budget and computing capability that should be tightly optimized for given tasks. In view of this, this paper aims to characterize promising use cases of critical mMTC and search for their possible solutions. To this end, we first review the state-of-the-art (SOTA) technologies for separate mMTC and URLLC services and then identify key challenges from conflicting SOTA requirements, followed by potential approaches to prospective critical mMTC solutions at different layers.

## I. INTRODUCTION

With the growing presence of enhanced wireless networks, ubiquitous infrastructure and vertical Internet of Things (IoT) domains (e.g., industry automation, autonomous vehicles), the evolution towards a comprehensive enabling platform for intelligent and tightly connected societies has already been evident. Such a platform could potentially consolidate the ever-increasing IoT applications, smart factory processes, public and businesses connectives jointly to uplift the overall quality of our daily lives. There are competing advances in both areas: i) licensed cellular systems, such as narrow-band IoT (NB-IoT) and long-term evolution for machine-type communications (LTE-M), and ii) unlicensed bands such as LoRa and Zigbee, to support the connectivity of the vertical IoT domains. We focus on cellular IoT in this paper as cellular systems have been celebrated due to their intrinsic benefits, including the provision of better quality of service and reliable coordination. The ITU radiocommunication sector (ITU-R) has cate- gorized 5G services into three broad classes, viz. mas-

S. R. Pokhrel, J. Ding, J. Park and J. Choi are with the School of Information Technology, Deakin University, Geelong, VIC 3220, Australia. Ok-S. Park is with Electronics and Telecommunications Research Institute, Daejeon, South Korea. A new version appears in IEEE Access [1] sive machine type communications (mMTC), ultra-reliable low-latency communications (URLLC) and enhanced mobile broadband (eMBB). A simplified view of the identified goals for the third generation partnership project (3GPP) fifth gen- eration (5G) can be outlined as follows:

- yield up to 20 Gbps of eMBB speed and increase sub- scriber capacity by about hundred times of the current levels. We may not see this speed in practice, but we should see the downlink speeds uplifting from 100 Mbps to Gbps.
- deliver URLLC, i.e., > 99.999% block error rate (BLER) and <1 milliseconds (ms) for some vertical IoT domains.
- enable massive connectivity, mMTC implementation and incorporation of low-power wide area network (LPWAN) requirements to strengthen IoT solutions.

These are all challenging goals, but have been mostly accom- plished. Latency is still a concern in the optimal present levels of 10 milliseconds to 30 milliseconds range. The reliability and latency requirements for seamless operations of critical emerging mMTC applications (e.g., the advanced driving-assistance technologies for vehicles in a massive smart city application, their autonomous cars and loss/delay sensitive industry 4.0 MTC for large factory automation) can barely be compromised. Therefore, our primary goal in this paper is to present a comprehensive review to summarize the current state of the art (SOTA) for URLLC within mMTC, refer to as critical mMTC, which is poorly understood in the literature.

To this end, one of the representative vertical domains is cellular-based smart manufacturing, which is characterized by the integration of IoT and related services in industrial automation. Standardized wireless network technologies and transmission protocols including 5G new radio (NR) are ineffective or unable to support low delay and ultra-high reliability requirements of such emerging critical mMTC ap- plications [2]. Intelligent integration of URLLC and mMTC has the potential to be a key enabler for the cellular-based industrial automation. As summarized in Table I, in many such not-so-futuristic applications areas like i) augmented and virtual reality, ii) future industrial communication and control, iii) massive network of autonomous vehicles and sensors, iv) haptics, robotics, and tactile Internet in large factory networks; the utilization of extremely reliable and virtually zero delay wireless communications are important and indispensable along with the availability of massive connectivity [3], [4]. New standardization and design guidelines for critical mMTC services should be developed to fulfill their virtually error free and zero delay constraints.

| Class | Reliability % | Delay ms | mMTC Apps. (see below) |
|---|---|---|---|
| 1 | $99.9 - 99.99999$ | > 50 | industry 4.0, vehicles, Haptics |
| 2 | $99.9 - 99.99999$ | 10 – 50 | IIoT automation/orchestration |
| 3 | 99.9 - 99.999 | 2 – 10 | Vehicles, AR/VR, Drones |
| 4 | 99.99999 | 2 | Autonomous Cars, Haptics |
| 5 | 99.999 | 1 | Vehicles, Haptics |
| 6 | 99.99999 | 2 | Vehicles, Internet of Drones |
| 7 | 99.999 | 2 | AR/VR, Convoy of Drones |
| 8 | 99.999 | 0.5 | Cars, AR/VR Drones |

| Applications | Cases | Delay in ms | Reliability |
|---|---|---|---|
| Virtual Reality | Haptics | 0.5 – 2 | 99.9 |
| | | | > 99.999 |
| Augmented Reality | Video | 0.5 – 2 | > 99.999 |
| Immersive Reality | 3D-Audio | 0.5 – 2 | 99.9 |
| Industry 4.0 | Haptics | highly-dynamic: 0.5 dynamic: 5 fixed: 50 | 99.9 > 99.999 |
| | Video | 2 | > 99.999 |
| | Audio | 2 | 99.9 |
| Autonomous Vehicles | Haptics | life critical: 0.5 | 99.9 |
| | Sensor | dynamic: 5 | > 99.999 |
| | Video | dynamic: 5, fixed: 50 | > 99.999 |
| | Audio | | 99.9 |
| Convoy/Swarm of Drones | Haptics | kinesthetic: 0.5 Tactile: 5 | 99.9 > 99.999 |
| | Sensor | 5 | > 99.999 |
| | Video | dynamic: 5, fixed: 50 | > 99.999 |
| | Audio | 1-5 | 99.9 |
| | GPS | 5 | 99.9 |
| Haptics communication (human touch feel) | Haptics | interaction: 0.5 Observation: 5 | 99.9 > 99.999 |
| | Video | dynamic: 1, fixed: 5 | > 99.999 |
| | Audio | 1-5 | 99.9 |

TABLE I: Reliability and latency requirements of a few mMTC applications [10]–[12].

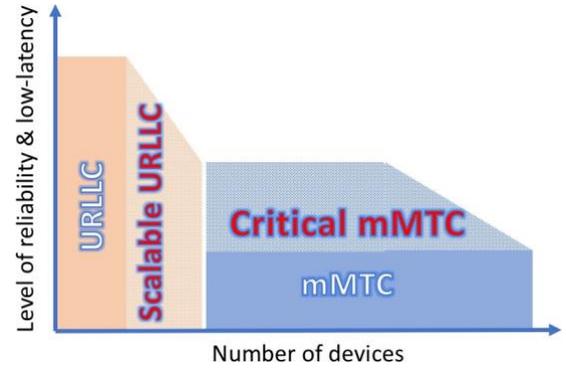

Fig. 1: Future Cellular IoT Services

In summary, both mMTC and URLLC need to set key parameters for higher layers to support heterogeneous emerging applications and services in the 5G systems, pursuing scalability and high reliability with low-latency, respectively. URLLC and mMTC are now envisaged to agglomerate together into critical mMTC, producing new challenges in designing systems beyond 5G. The factory automation application, for example, has evolved from wired massive sensors and actuators towards wireless intelligent machines and sensors for enabling flexible and smart manufacturing. A simple mixture of conventional mMTC and URLLC (e.g., radio access network slicing [5], [6]) cannot simultaneously guarantee their reliability, latency, and scalability over wireless links at the same level as perceived form wired factory environment (e.g., < 4 ms latency, $10_6$ devices/km²). Furthermore, generic solutions for scalable URLLC (e.g., machine learning (ML) aided URLLC for autonomous vehicles [7]) are ill-suited to critical mMTC [8], [9].

### A. Motivation of Critical mMTC

URLLC and mMTC are the two key service types supported by 5G, achieving reliability with low latency and scalability, respectively. As shown in Fig. 1, it is envisaged that these two services will agglomerate into critical mMTC or scalable URLLC for beyond 5G with emerging use cases. In this subsection, we aim to justify the need of critical mMTC by highlighting key differenes from existing similar notions.

○ URLLC-mMTC Mixture is the use case where URLLC and mMTC services are supported with separate physical resources by the same network. To support this heterogeneous system, the cellular resource configuration (i.e., numerology) has been re-designed [13], [14], and the mini-slot resources therein have been optimized using radio access network (RAN) slicing methods in an orthogonal or non-orthogonal resource allocation [6].

○ Scalable URLLC is targeted to support increased connections of MTC devices with a variety of URLLC requirements. This traffic type is non-separable, and thus cannot be supported using neither RAN slicing nor prioritizing methods used for the URLLC-mMTC mixture. Instead, ML based solutions have recently been proposed [7] in which prediction is adopted for reducing latency and improving reliability, without consuming radio resources.

○ Critical mMTC is targeted to support enhanced URLLC requirements for a fraction of massively connected MTC devices. It is possible to apply some techniques developed for the URLLC-mMTC mixture to critical mMTC. However, when a device can sometimes be a URLLC or non-URLLC device, resource configuration overhead may be overwhelming. Notwithstanding, it cannot be the same, especially for a large number of critical MTC devices that apply to scalable URLLC. Meanwhile, MTC devices generally lack powerful computing power and sufficient energy, so ML-based solutions for scalable URLLC are not suitable unless lightweight and energy-efficient ML architectures and algorithms are deployed.

In brief, critical mMTC is still a unique service type that has not been fully understood. This mandates our seeking for novel solutions to critical mMTC by unveiling its key characteristics and rethinking existing relevant approaches.

### B. Review of Related Survey

Of particular relevance to this work is the survey and overview research works of URLLC and mMTC for 5G IoT. Table II provides a summary of references related to these areas categorized based on their themes. Several researchers have provided a review of enabling approaches, ideas, challenges and applications of URLLC/mMTC in different IoT contexts [2], [4], [15]–[27], but the area of critical mMTC has been poorly reviewed in literature [17].

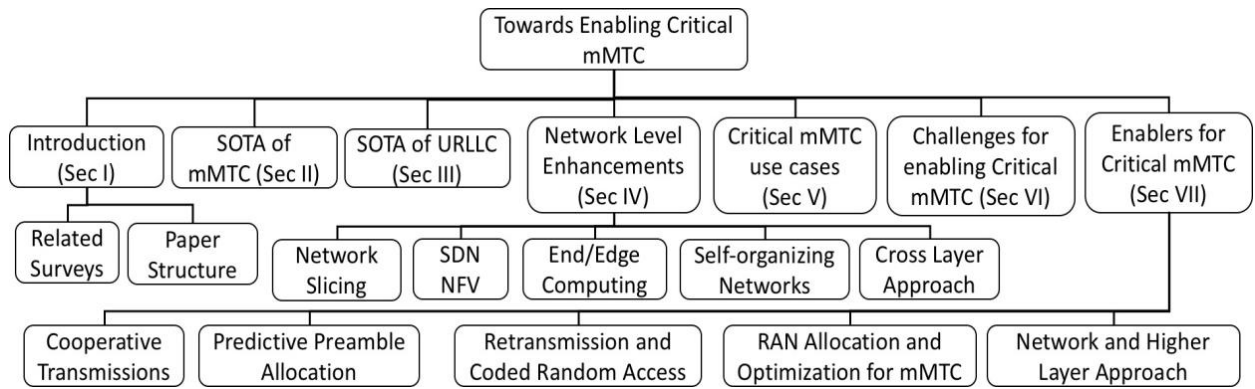

Fig. 2: A high level view of topics and subtopics of the paper

Furthermore, researchers have already summarized RA based schemes for the URLLC [28]–[31] and mMTC [31], [32] separately. A few other researches have covered a survey of the existing machine learning [36]–[38] and data analytics [33]–[35] approaches for URLLC and mMTC. A couple of surveys [39], [40] of highly densed IoT are relevant to the mMTC cellular IoT framework. Nevertheless, a review of enablers, challenges and techniques for the convergence of critical mMTC, our main theme in this paper, is a challenging and delicate task because of the complexity due to the heterogeneity in service and application constraints discussed earlier (recall Table I).

### C. Contributions

While some of the recent survey/vision papers (see Tab. II) considered various facets of mMTC and URLLC schemes, a thorough study of the research problems involved in supporting the large number of devices with a mix of mMTC and URLLC services in cellular IoT networks and a systematic evaluation of recent advances in 5G to resolve the challenges is missing in literature. As mentioned earlier, several challenges arise while incorporating URLLC and mMTC together into existing NBIoT-based cellular systems. Key concerns include the provisioning of QoS to heterogeneous mMTC systems, for example, fixing URLLC within mMTC, QoS-aware transmission scheduling, RAN congestion avoidance etc.

Overall, we highlight the main contributions of this paper as follows:

- Major challenges faced by the current IoT (cellular) networks to support URLLC within a large number of mMTC devices are classified and possible future innovative ideas are noted along with essential elements, traffic characterization and potential vertical application domains.

- Few ineffectiveness of the standard mMTC guidelines are acknowledged. With their application to critical mMTC systems, new ideas were proposed along with key features and channel control procedures for the evolving Cellular IoT standards.
- Current higher-layer approaches for handling URLLC and mMTC separately in cellular IoT networks are reviewed along with insights on new concepts for overcoming critical mMTC specifications (jointly handling URLLC and mMTC) from higher-level perspectives.
- Multiple research problems are identified and also some promising potential directions to promote future research efforts in the relevant fields are discussed.

### D. Structure of the Paper

To improve the readability of this paper, we provide a high level view of the paper structure in Fig. 2 and the definitions of acronyms in Table III. The remainder of the paper is structured as follows. Section II summarizes the SOTA of mMTC along with a discussion on NB-IoT, RAN and core network enhancements provisioning for massive connectivity. Section II also highlights some open issues for enabling massive connectivity. Section III reviews SOTA of URLLC with basic functionalities. Subsequently, Section IV highlights network level enhancements for both URLLC and mMTC. Section V presents important use cases of critical mMTC with some existing solutions. Section IV provides challenges towards enabling critical mMTC along with the difficulties at the network level. Finally, Section VII provides future directions towards potential critical mMTC enablers, and Section VIII concludes this paper.

## II. STATE-OF-THE-ART OF mMTC

Most of mMTC use cases are characterized by a scenario where a large number of machine-type devices deployed in a wide area sporadically communicate without specific latency requirements, ignoring reliability guarantees. Supporting these use cases is commonly recast by meeting the following key performance indicators (KPIs) [41]:

- Massive connection density of $10_6$ devices per square kilometer in an urban environment;

| References | URLLC | mMTC | Scope |
|---|---|---|---|
| [2], [4], [9], [15]–[27] | [17]–[19], [22]–[27] | [2], [4], [16]–[21] | Enablers, challenges, applications |
| [28]–[32] | [28]–[31] | [31], [32] | Random Access (RA) |
| [33]–[35] | ✓ | ✓ | QoS, Data Analytics |
| [36]–[38] | ✓ | ✓ | Machine Learning |
| [39], [40] | – | ✓ | Highly Dense IoT |

TABLE II: Survey and overview works in the areas of 5G URLLC and mMTC.

| Abbreviations | Definitions |
|---|---|
| 3GPP | Third Generation Partnership Project |
| 4G | Fourth Generation |
| 5G | Fifth Generation |
| WUE | Wake Up Signal |
| BLER | Block Error Rate |
| BPSK | Binary Phase Shift Keying |
| BS | Base Station |
| CS | Compressive Sensing |
| DL | Downlink |
| DPSK | Differential Phase Shift Keying |
| DQPSK | Differential Quadrature Phase Shift Keying |
| FDMA | Frequency-Division Multiple Access |
| GSM | Global System for Mobile Communication |
| ISM | Industrial, Scientific, and Medical |
| ITU | International Telecommunication Union |
| IoT | Internet of Things |
| LPWAN | Low Power Wide Area Networks |
| LTE | Long-Term Evolution |
| LTE-A | Long-Term Evolution-Advance |
| LTE-M | Long-Term Evolution Machine Type Communications |
| ML | Machine Learning |
| MTC | Machine Type Communications |
| NB-IoT | Narrow-Band IoT |
| NFV | Network Function Virtualization |
| NOMA | Non-Orthogonal Multiple Access |
| OFDM | Orthogonal Frequency Division Multiplexing |
| OFDMA | Orthogonal Frequency Division Multiplexing Access |
| OOK | On-Off Keying |
| OQPSK | Offset Quadrature Phase-Shift Keying |
| PDM | Power-Domain Multiplexing |
| PUSCH | Physical Uplink Shared Channel |
| NPRACH | Narrowband Physical Random Access Channel |
| PRB | Physical Resource Block |
| PSM | Power Saving Mode |
| QAM | Quadrature Amplitude Modulation |
| QPSK | Quadrature Phase-Shift Keying |
| QoS | Quality of Services |
| RA | Random Access |
| RAN | Random Access Network |
| RF | Radio Frequency |
| RL | Reinforcement Learning |
| RRC | Radio Resource Control |
| SC-FDMA | Single-carrier Frequency-Division Multiple Access |
| SDN | Software Defined Networking |
| SIC | Successive Interference Cancellation |
| SINR | Signal-to-Noise-and-Interference Ratio |
| SLA | Service Level Agreement |
| TDD | Time Division Duplex |
| TDMA | Time-Division Multiple Access |
| TTL | Time-To-Live |
| TTI | Transmission Time Interval |
| UAV | Unmanned Aerial Vehicles |
| UL | Uplink |
| V2X | Vehicle-to-Everything |
| WLAN | Wireless Local Area Networks |
| WPAN | Wireless Personal Area Networks |
| WAC | Wide Area Coverage |
| WSN | Wireless Sensor Networks |
| mMIMO | Massive Multiple-Input Multiple Output |
| eDRX | Expanded Discontinuous Reception |

TABLE III: Definitions of abbreviations.

○ Maximum coupling loss (MCL) up to 164 dB for wide coverage; and

○ Device battery lifetime over 10 years with a stored energy capacity of 5 Wh.

To achieve these mMTC KPIs, 3GPP has developed two radio technology standards built on the existing long term evolution (LTE) systems: LTE-M for lightweight mobile ap-plications and voice-over-LTE (VoLTE); and NB-IoT for low-rate and wide-coverage applications. At ITU-R WP5D#32 meeting in July 2019, NB-IoT was officially recognized as a 5G candidate solution to meet the technical requirements of large-scale mMTC scenarios. Hereafter, we therefore focus on NB-IoT as a representative mMTC technology, and overview its key features as well as major enhancements.

*A. Features of NB-IoT*

NB-IoT is a radio technology proposed in 3GPP Release 13 towards supporting narrow-band LPWAN, a commercially successful wide-area mMTC application. Since NB-IoT is built upon LTE networks, NB-IoT can exploit LTE network hardware and resources, thereby reducing the deployment as well as operational costs [41]. At the physical layer, each NB-IoT carrier requires a minimum bandwidth of 180 kHz, which is equivalent to one LTE Physical Resource Block (PRB), and has three operational modes utilizing different amount of spectrum [42]. Specifically, NB-IoT can reserve: one or more LTE PRBs within an LTE carrier under the in-band mode; one or multiple global systems for mobile communications (GSM) carriers under the stand-alone mode; and only the spectrum within the guard-band of an LTE carrier under the guard-band mode.

*1) Massive Connectivity:* To achieve massive connectivity, resource unit (RU) configurations based on subcarriers or tones are introduced for multiple access [43]. Specifically, single-carrier frequency-division multiple access (SC-FDMA) is applied to uplink, using either 3.75 kHz or 15 kHz subcarrier spacing. With a 3.75 kHz subcarrier spacing, NB-IoT can only allocate single-tone RU with a duration of 32 ms to different users. This mode is designed for power-constrained devices in a wide area. In addition, the reduced subcarrier bandwidth allows the simultaneous allocation of up to 48 devices within one PRB. On the other hand, with a 15 kHz subcarrier spacing, either single-tone RU (8 ms) or multi-tone RUs (3, 6, or 12 tones) with a duration of 4 ms, 2 ms, and 1 ms, respectively, can be configured to different devices. For single-tone configurations, phase rotated $\pi/2$-binary phase shift keying (BPSK) or $\pi/4$-quadrature phase shift keying (QPSK) modulations can be used. For multi-tone configurations, only QPSK modulation is used [44]. For narrowband physical random access channel (NPRACH) of NB-IoT, a single-tone based preamble of 4 symbol groups with frequency hopping is designed [45]. In addition, as in LTE, orthogonal frequency-division multiple access (OFDMA) is used in the NB-IoT downlink with a 15 kHz subcarrier spacing only.

*2) Wide Coverage:* In order to extend the coverage range in an open environment and compensate the penetration losses in a challenging indoor space with highly reliable communication, NB-IoT targets up to 164 dB MCL, which is 20 dB coverage enhancement compared to GSM and general packet radio service (GPRS). To achieve this goal, except operating in narrow bandwidth, NB-IoT supports the approach of transmission repetitions (i.e., exploiting time diversity), by which the received signal-to-noise ratio (SNR) can be enhanced so that data could be decoded even when the signal power is much lower than the noise power [46]. In NB-IoT, the signal transmission can be repeated up to 128 and 2048 times in the uplink and downlink, respectively, and the number of repetitions depends on the coverage enhancement level required by devices along with the number of tones and subcarrier spacing [47].

*3) Low Power Consumption:* To lengthen battery life of devices, wake up signal (WUS) is used in NB-IoT to allow devices to avoid regularly paging checking and only start the procedure once WUS is received [48]. In addition, NB-IoT inherits two power saving techniques of LTE, i.e., power saving mode (PSM) and expanded discontinuous reception (eDRX) [49]. In particular, PSM allows a device registered on a network to turn off the functionalities of paging listening and link quality measurements to save energy. With eDRX, a device can negotiate with a network the time that it can fully turn off the receiving functionality for energy saving.

### B. Enhancements for NB-IoT

The continued evolution of NB-IoT is an important 3GPP activity. For instance, 3GPP Release 14 had enhancements in the form of support for higher data rates, multicast, positioning, a lower power class, and system access on non-anchor carriers [19]. 3GPP Release 15 included further enhancements in the form of support for improved latency, power consumption, measurement accuracy, cell range and load control. To extend the range of deployment options, it also specified small-cell and time division duplex (TDD) support for NB-IoT [48]. In 3GPP Release 16, additional enhancements and extensions for NB-IoT are specified to further improve the efficiency of network operation. Particularly, improvements on spectral efficiency for NB-IoT transmission and energy efficiency for NB-IoT devices are studied and proposed [13], which include: 1) enhanced mobile-terminated early-data transmission; 2) support for device-group WUS; and 3) improved uplink transmission using pre-configured resources in idle mode. Furthermore, 3GPP Release 16 also includes a common 2-step PRACH to decrease the latency and reduce additional control-signaling overhead. This is achieved by combining the preamble and the scheduled physical uplink shared channel (PUSCH) transmission into a single message from devices, known as MsgA. Then by combining the random-access respond and the contention resolution message into a single message (MsgB) from the BS to devices [13].

In addition to the enhancements by 3GPP, it has been conducted to analyze and improve the performance of NB-IoT networks in a range of areas including: 1) NPRACH evaluation and enhancements [50]–[54]; 2) coverage analysis and enhancements [55]–[58]; 3) energy efficiency enhancements [59]–[61]; and 4) co-channel interference analysis and mitigation [62]–[65].

### C. RAN & Network Enhancements for Massive Connectivity

3GPP has continued to expand the 5G infrastructure with new mMTC technologies for cellular IoT support, 5G LAN services, industry 4.0 time-aided networking and optimized access and relay backhaul [11]. It has also developed features for improving service-oriented architecture, enhancing flexible session management deployments, user plane and control functions, and supporting commercial services using location-based service architecture, enhancing self-organizing networks, smart dual connectivity, and aggregating dynamic carrier [11, Sec 4.2].[1]

*1) 5G enhanced support for Vertical IoT Domains:* Some of the key features identified by 3GPP relevant to our context are: enhanced support of LAN services (Vertical LAN), 5G support for Cellular IoT evolution (5G CIoT), support for Industrial IoT (NR IIoT), architectural enhancements for advanced vehicle to everything (V2X) services (eV2XARC) and enablers for network automation and orchestrations [18], [27].

*2) Service based architecture enhancements:* Service based enhancements scale and help for flexible implementation of proxy-based infrastructure and communication network through repository function and service [66]. The network functions and services defined can be implemented and used selectively by maintaining a set of services and a set of functionalities.

*3) Enhanced network and RAN slicing:* The network slicing provides a mechanism for reallocation of access and mobility management along with session management and control functions [18] . It also facilitates independent and slice specific authentication and authorization per network slice. Starting from the granularity of network slicing per user or one slice for each services, we can identify how these slices are mapped and implemented at the RAN level in an efficient way in terms of radio resource consumption. We can start from the lower layers up to network function selection, configuration and chaining for each slice [67].

*4) Enhanced Automation and Orchestration:* There are advancements in the data collection and network analytic features specific to data collection based on the source and type of information to network analytics of mMTC, their automation and orchestration functions such as slice specific load balancing, patching, evaluation of network performance indicators, user mobility and sustainable QoS [68]. With such enhancements, the network is capable of utilizing multiple data analytics, more importantly, analytics support are available at several levels.

*5) Architecture support for time-based networking:* New 5G architectural enhancements enable services to provide time synchronization of packet delivery in each hop of mMTC, to support time sensitive networking such as industrial automation [69]. 5G system can be easily integrated with an external network for desired level of services [70]. For example, a centralized model and with a specific subset of specifications and features can be enabled for such integration. Typical areas include periodic QoS guarantees, store and forward buffer, and logical bridging management for desired QoS mapping. Such a time-based approach includes a unique and even multiple working domains via a single RAN architecture [11].

---

[1]https://www.3gpp.org/news-events/2122-tsn_v_lan.

## III. State-of-the-art of 5G URLLC

URLLC aims to support mission-critical applications ranging from factory automation to vehicle safety control. In sharp contrast to conventional communication services pursuing the best-effort performance, these mission-critical applications commonly strictly require high reliability and low latency guarantees (e.g., $> 99.999\%$ BLER with $< 1$ ms user-plane latency [13]). In 3GPP specifications, new radio (NR) standards have been developed to enable URLLC. In this section, we provide an overview of key URLLC features and major enhancements in NR.

### A. Basic URLLC Functionality

Most of the basic URLLC functionalities in NR were filed within 3GPP Release 15 and finalized in September 2018. Thus, we focus on 3GPP Release 15, and elaborate the key functionalities provisioning URLLC service level agreement (SLA) in terms of latency and reliability [13].

*1) Functionality for Low Latency:* Two main functionalities for low latency are explained as follows.

*a) Flexible Numerology and Frame Structure:* NR supports scalable numerology and flexible framework to address heterogeneous configurations, deployment, and services [71]. In particular, subcarrier spacing of $2^n \times 15$ KHz ($n = 0, 1, 2$) can be supported for data channels at sub-6 GHz in NR. To achieve the low latency target of URLLC, it is necessary to use short transmission time interval (TTI). In light of this, large subcarrier spacing should be employed for URLLC transmission. For example, the TTI of 14 OFDM symbols with 15 kHz subcarrier spacing is 1 ms, while that with 60 kHz subcarrier spacing is reduced to 0.25 ms. Furthermore, NR also introduced the concept of mini-slot (consisting of only 2, 4, or 7 OFDM symbols) to further shorten TTI. For example, the TTI of a 2-symbol mini-slot with 60 kHz subcarrier spacing is only 35.7 μs, which facilitates to meet the low-latency goal.

In addition, other flexible framework designs such as frequent transmission opportunities, self-contained slot structure, shortening hybrid automatic repeat request (HARQ) roundtrip time, and traffic preemption for URLLC are also the important features to enable low-latency transmissions [10].

*b) Configured-Grant (Grant-Free) scheduling:* The conventional grant-based PRACH requires each device to transmit a scheduling request first and then wait for an uplink grant from the BS. The complex handshaking procedure between devices and BS can result in excessive signaling overhead and makes it difficult to meet the URLLC latency requirement of 1 ms or less in some scenarios. To reduce the access delay, configured-grant (grant-free) scheduling in 3GPP was proposed to allow devices to transmit uplink data without uplink grant on the configured resources [72]. The periodicity of configured resources can be as short as 2 symbols. The devices in grant-free scheduling can be dedicated with periodic traffic for non-contention based PRACH or a group of devices with aperiodic or sporadic traffic for contention based PRACH.

*2) Functionality for High Reliability:* Three important functionalities for high reliability are discussed in the following.

*a) Low Spectrum-Efficient MCS/CQI:* To guarantee reliable data transmission, optimal modulation and coding scheme (MCS) based on the channel quality indication (CQI) should be employed according to a look-up table [73]. In URLLC, reliability can be improved at the expense of low spectrum efficiency. To this end, one can either enlarge resource bandwidth or shorten TTI, which is of help to use low coding rate for reliability enhancement. Therefore, low spectral efficiency entries, e.g., QPSK with 1/8 coding rate, are considered in MCS table for URLLC.

*b) Multi-Slot Repetition:* Repetition is a common approach to improve reliability as well as coverage. In URLLC, multi-slot repetition (2, 4, and 8 repetitions) can be used in data channels without waiting for any grant or retransmission feedback, which is also useful in URLLC for devices without sufficient time to provide or wait for HARQ feedback. Note that repetitions can also be applied to the control channels.

*c) Diversity Exploitation:* In URLLC, time diversity is not a viable solution for reliability enhancement as the packet cannot span over a long time due to a tight latency budget [74]. However, exploiting diversity in the frequency and spatial domains could be the ways to improve the reliability of URLLC. For example, frequency diversity could be achieved by frequency hopping and spatial diversity can be achieved by dual-connectivity or multi-connectivity of non-collocated BSs, or by multiple-input multiple-output (MIMO or massive MIMO) technologies.

### B. Enhancements for URLLC

3GPP Release 15 was revolutionary in terms of introducing a brand-new NR for URLLC, whereas Release 16 and 17 were evolutionary, targeting new verticals with tighter requirements by enhancing the capacity and operation of existing features [13]. In particular, Release 16 has expanded URLLC to the new verticals such as factory automation, transport industry, and electrical power distribution. To support more strict URLLC requirements, it focused on the reliability enhancements on the control message transmission and latency reduction on HARQ feedback. Moreover, it studied the potential benefits of uplink inter device prioritization and multiplexing including uplink preemption and enhanced power control. It also supported multiple active configurations for configured-grant enhancements, so that different service traffic can be accommodated and quicker alignment for URLLC uplink transmissions can be made. In addition, the 2-step PRACH proposed in Release 16 can also be applied to URLLC. Release 17 will bring more new use cases to enable everything connected for 5G evolution systems in 2020 and on wards. Particularly, studies of URLLC in Release 17 will mainly focus on new emerging verticals and end-to-end performance of different applications, building enhancements to Release 16 features [72].

In addition to the studies and enhancements by 3GPP, several research works have also contributed to the analysis and enhancements for URLLC. For instance, in [75], a tractable approach is proposed to derive and analyze the latent access failure probability of URLLC device under three different grant-free RA schemes, namely reactive, K-repetition, and proactive, and showed that the proactive scheme can provide the lowest latent access failure probability under shorter latency constraints. In [76], the shortcomings of existing grant-free RA schemes are discussed for enabling URLLC, and two advanced grant-free RA schemes that go beyond 5G NR are proposed by taking advantage of non-orthogonal multiple access (NOMA). Furthermore, in [77], repetition for NOMA has been studied to lower outage probability.

## IV. Network Level URLLC and mMTC Enhancements

While much attention has been focused on the physical and link layers, it is increasingly being realized that a wider redesign at network level is also essential to meet the specified requirements. It is also known that the existing cellular network architecture cannot support diversified services as network design fundamentals remains unchanged – largely based on conventional mobile broadband services. Thus, novel mobile computing frameworks and flexible network architectures are to be developed. These include approaches such as network slicing, software-defined networking (SDN), network function virtualization (NFV), orchestration and self-organizing networks (SONs). In this section, we cover each of these approaches and their advancements for URLLC services.

### A. Dynamic and Constrained Network Slicing

Network slicing has been a fundamental technique to harness diversified services simultaneously over the same physical infrastructure for future networks by enabling the deployment of multiple virtual domains atop a shared infrastructure [5], [6]. In scenarios like eMBB, mMTC, and URLLC, network slicing not only i) allows building multiple logical subnetworks with reserved resources enhancing the quality of service but also ii) avoids interruptions caused by other services. For example, to accommodate a number of critical mMTC (different) scenarios, or URLLC within mMTC use cases, we should be able to manipulate the network on the fly. For this vision to be realistic, network operators should be able to orchestrate their capabilities across several points in the infrastructure dynamically – dynamic network slicing [78], [79]. Furthermore, to satisfy critical mMTC requirements, we need a different approach to resource allocation among network slices for guaranteed and context aware slicing of resources– constrained network slicing [80].

The primary benefit to exploit such a dynamic network slicing for critical mMTC is that the network operators will then be able to plug and play with the slices timely to satisfy both URLLC and mMTC temporal services. For example, constrained dynamic network slicing can be used to manage the time varying network traffic, when an event occurs or disaster happens, the area get overwhelms with downlinks first as most devices download information about the event before or when it begins. However, after sometime, it will be opposite in the sense that the whole traffic shifts to the uplink because devices then start streaming live and uploading data/information about the event to the server for data analytics to generate next actions. To attain such a constraint and dynamic network slicing, the slices should be optimizing their resource allocations under loss and delay constraints and satisfy the URLLC requirements [81]. In particular, constrained dynamic optimization may help to guarantee that the slices are managed in a desired fashion and the network resources are shared to meet their SLAs depending on the applications' need at the higher layers.

More importantly, such slicing would be able to tailor the infrastructure by observing the time-evolving traffic pattern and network dynamics. As mentioned, network slicing has the potential to tackle and capitalize such scenario by utilizing the capabilities of NFV, orchestration, SDN and analytics [82]. However, strictly isolating and/or dynamic sharing of resources among multiple slices has been facing significant challenges because of the corresponding traffic fluctuations due to evolving users and varying channel conditions, therefore, it requires further investigations [6], [83]. Other intelligent solutions utilizing data-driven machine learning and artificial intelligence become crucial for several vertical domains including but not limited to, more efficient manufacturing, urban computing and autonomous traffic settings [84].

### B. Software Defined Networking and Network Function Virtualization

SDN and NFV are the two successful network architecture techniques that support dynamic network slicing [70], [85]. Caballero et al. [85] illustrated via game-theoretic tools that simplification of scheduling and resource allocation in software-defined networking can be accomplished by separating the network control plane from the packet forwarding plane which offers dynamic network flow management. Additionally, network function visualization provides a high degree of programmability and flexibility by decoupling network functions from dedicated hardware devices. We have observed that integrating these two techniques can significantly provide efficient, scalable, and flexible network slicing service configurations for improving the overall performance for the network layer. In fact, it ameliorates both latency and reliability substantially [18]. It has been reported that software-defined networks can achieve an up to 75% performance improvement in end-to-end latency [3]. Ksentini et al. [86] proposed two-level medium-access-control scheduling framework for slicing by using dynamic slice management and established the QoS requirement for URLLC and mMTC. Zheng et al. [87] proposed constraint enabled resource slicing.

### C. Edge and End Node Computing

Mobile edge/end node computing is another best solution that can reduce latency while processing tasks in URLLC with massive connectivity [81]. Mobile edge computing can be integrated with software-defined networks and network function virtualization to deal with such service disruption [88]. Improvement in resiliency and reduction in latency can be significantly observed when distributed, and virtualized networks are provisioned efficiently. Computation and transmit power can be significantly saved by optimizing the offloading process and required resource allocation in such mobile edge computing frameworks. However, the tradeoff between delay and computing power requires further investigations along the lines of [89]. Sood et al. [90] proposed end node computing and reported several challenges. Further extensive investigations are required to determine the best optimization techniques at the edge and end devices from the task offloading and scheduling prospective to provide a high quality of service in the URLLC of massive connectivity.

### D. Self-organizing Networks

All of the aforementioned techniques, like network slicing, software-defined networks, and network function visualization improve network scalability and flexibility, however, the quality of service and perceived user experience may be compromised due to the integration of the techniques and complication in network management and their parameter configuration. Therefore, it is critical to adopt self-organizing network management strategies, which could provide the desired optimization, distributed management, intelligence, and automation in such convergent network layer approaches [91]. In [92], a joint scheduling framework is proposed with traffic steering for eMBB and URLLC. However, to attain these, an autonomous intelligent service-aware SON is required for the such network with control functions at a network layer central unit, and various edge and core servers [93]. A new catalog-driven network management system that enables the smart deployment of service has already been proposed by the European Union's 5G Public-Private Partnership Project. Such self organization can be achieved by several machine learning techniques as outlined in [94].

### E. Cross-Layer Enhancements

Cross-layer design has the potential to improve both the end-to-end delay and the overall reliability significantly. This is feasible as each layer of the TCP/IP protocol stack is inherent and interdependent with the other layers. As an example, the transmission of packets, queuing in buffers, and delays in routing mechanisms all depend on the dynamics of the physical, link, and network layers, respectively. Better resource utilization can definitely be achieved by optimizing the delay components based on the end-to-end delay constraints. To this end, the investigation for bandwidth and energy optimization by adopting a cross-layer design is on real momentum [95], [96].

Collins and Cruz [97] introduced, first of all, the concept of cross-layer scheduling on the basis of the queueing and channel conditions in order to minimize average power usage under the average delay limit. Wang et al. [98] developed a cross-layer approach to manage data networking latency in wireless sensor networks, resulting in better energy efficiency.

She et al. [99] adopted the cross-layer technique in-radio communication network for URLLCs (given delay is shorter than the width of the signal). More importantly, authors in [95], [96] studied cross-design approach for offloading and NOMA systems.

## V. CRITICAL mMTC USE CASES AND EXISTING SOLUTIONS

Based on the overviews of mMTC and URLLC in the preceding sections, we summarize the SLAs of mMTC and URLLC in Table IV.

| Specifications | mMTC | URLLC |
|---|---|---|
| Connection Density | Up to $10_6$/km$^2$ | Comparably low |
| Power Consumption | Extreme low | Insensitive |
| End to End Latency | Insensitive | $1 - 10$ ms |
| Reliability | Typical BLER $10_{-1}$ | Up to BLER $10_{-9}$ |
| Payload Size | Small | Small to large |
| Bandwidth | Narrowband | Wideband |
| Numerology | 3.75 KHz 15KHz | $2_n \times$ 15KHz |

TABLE IV: Summary of Requirement and Specification Disparity between mMTC and URLLC

Given the distinct SLAs of mMTC and URLLC, in the this section we focus on possible critical mMTC uses cases wherein a massive number of non-critical MTC and critical MTC devices coexist, and discuss existing solutions to such use cases.

### A. Critical mMTC Use Cases

Critical MTC use cases, requirements, and traffic characteristics have been discussed in [10]–[12], as summarized in Table V. These critical MTC devices are expected to coexist with non-critical MTC devices, forming the heterogeneous traffic of critical mMTC.

One exemplary use case of critical mMTC is studied in [100] from smart city perspective. In the smart city, a massive number of devices perform various tasks ranging from environmental and critical infrastructure monitoring to smart grids and industrial automation. When a triggering event occurs, a fraction of these devices become sensing devices that observe critical triggering information. These devices are grouped based on their functionalities and/or geographic locations, forming a critical MTC group while the remainder becomes non-grouped mMTC devices. A critical MTC group header is then assigned for measuring the urgency level of each event and communicating with a BS through URLLC.

Fitzgerald et al. [101] studied another critical mMTC use case is investigated in the context of industrial automation and the use of massive MIMO. In this scenario, the traffic is categorized into control and alarm traffic classes. Control traffic encompasses the transmissions between machines on the factory floor and their controllers. This traffic is regular (even deterministic for some cases), and has stringent latency requirements in order to arrive within a specified control loop period. By contrast, alarm traffic is sporadic and unpredictable, but nonetheless must be delivered reliably. Each triggered alarm has its strict deadline, until which the alarm can be retransmitted to reach the BS. The allocation of massive MIMO spatial resources can be optimized for supporting such heterogeneous control and alarm MTC traffic.

| Use case | Reliability (%) | e2e Latency | Data packet size | Traffic |
|---|---|---|---|---|
| Smart city: (smart metering, waste management, and smart parking) | 99 | $\leq 5$ s | Mostly $\leq 200$ bytes | Sporadic |
| Public safety | 99.999 | 1 ms | Small to big | Sporadic |
| Factory automation (motion control) | 99.9999 | 2 ms | 32 bytes | Periodic deterministic |
| Remote driving | 99.999 | 5 ms | UL: 2.5 Mpbs; Packet size 5220 bytes; DL: 1Mbps; Packet size 2083 bytes | Periodic |
| Intelligent transport system | 99.999 | 10 ms | UL&DL: 1.1 Mbps; Packet size 1370 bytes | Periodic |

TABLE V: Critical MTC use cases, requirements, and traffic characteristics [10]–[12].

Thota et al. [102] simulates a mixed traffic of 5% URLLC and 95% non-URLLC devices whose arrivals are modeled by beta and uniform distributions, respectively. To enable the guaranteed access of URLLC devices, a number of preambles can be reserved. Under limited resources, however, this solution may result in a high collision probability of non-URLLC devices. Thus, to increase the preamble resource utilization of URLLC by exploiting the traffic characteristics (e.g., sporadic URLLC arrivals) is one key enabler.

### B. Existing Solutions

To support diverse 5G service types, several network slicing and edge computing solutions have been proposed. These existing solutions are also relevant to supporting critical mMTC, which we will briefly review in this section.

*1) Radio Access Network Slicing:* It is generally accepted that network slicing will address various use cases and networking capabilities of the emerging 5 G networks. Traditionally, radio access network (RAN) slicing has focused mostly on high layers (e.g., numerology, core network coordination issues in transport and application layers) while addressing their functionality split, computing resource allocation, and coordination. Recently, RAN slicing has also been focusing on radio resource allocation issues in lower layers (e.g., multiple access schemes in MAC layer, interference issues in PHY layer) [5], [6]. Resource allocation is vital for enhancing resource-multiplexing gain between slices while meeting specific RAN slicing service requirements. Regrettably, resource management in RAN slicing is a nontrivial task due to performance isolation, diversified service requirements and network complexities (including mobility and channel statuses). Nonetheless, for 5 G RAN slicing, one can build an intelligent resource allocation strategy. The main advantage of scheduling is to reap the benefits of a collaborative learning framework that comprises of deep learning in conjunction with reinforcement learning. For this function, deep learning can be used for large time-scale allocation of resources, while reinforcement learning can be used for online resource management to resolve small-time network dynamics, including imprecision projections and unpredictable network conditions.

*2) Low Latency Edge and End Node Computing:* Edge computing has been an established paradigm for distributed computing and control services by shifting network load from centralized server towards the last mile networks. At the heart of the 5G wireless systems and beyond is the edge computing capabilities at the cellular BSs. While existing state-of-the-art networks connect and process data centrally (in the cloud or at the edge) for delay and computation-centric applications, both wireless connectivity and database capacity need to be taken closer to the end nodes, thus raising the availability of storage and processing-enabled small cell BSs at the end nodes (e.g., vehicles with BS functionalities [103]).

In addition, the network infrastructure must provide a distributed decision-making service that understands and evolves to cope with the network dynamics to shorten delay. It optimizes communication networks and operations appropriately. One can provide a fresh look to the concept of end node computing by first discussing the applications that the network edge must provide, with a special emphasis on the ensuing challenges in enabling URLLC edge computing services for critical mMTC applications such as VR, V2X, haptics and so forth [81], [104]. Such edge and end node computing paradigm can be integrated with other features (such as software-defined SDNs and NFV) to cope with service disruptions [88], while improving resiliency and reducing latency.

*3) Service Aware Self-organizing Networks:* While the aforementioned techniques, including RAN slicing, edge and end node computing, and their integrations with SDN and NFV, improve the scalability and flexibility to support diverse service types, it may need to compromise the quality of service and perceived user experiences, due to the integration of multiple techniques and the complications in network configuration management. It is therefore crucial to adopt complementary self-organizing network management strategies, which could provide the desired optimization, distributed management, intelligence, and automation in such convergent network layer approaches [91].

## VI. CHALLENGES TOWARDS ENABLING CRITICAL MMTC

The main challenges involved in enabling critical mMTC in future cellular networks is shown in Fig. 3 and explained in the following.

### A. Mixed-Numerology Interference

Due to the contrasting requirements of URLLC and mMTC, service configuration differs significantly from the physical layer perspective [14]. In particular, mMTC usually characterized by narrowband transmission with small subcarrier spacing and low baseband sampling rate to support massive connectivity and wide coverage with low-cost and low-power consumption. On the other hand, URLLC usually requires a large subcarrier spacing to meet stringent latency requirement and has high baseband sampling rate. These heterogeneous configuration disparities in baseband and RF inevitably lead to significant interference [105], [106] in critical mMTC.

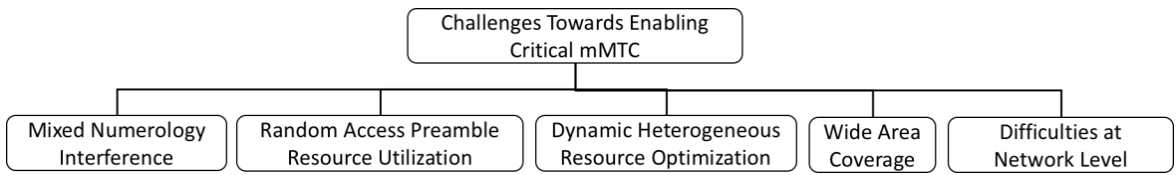

Fig. 3: Challenges of existing 5G to support critical mMTC communications.

### B. RA Preamble Resource Utilization

In 3GPP, grant-free RA or 2-step RA has been proposed and extensively discussed for both URLLC and mMTC, where the traditional grant request by devices prior to uplink transmission is omitted to reduce the control-signaling overhead and access latency. In grant-free RA, preamble resources are precious and scarce, which needs to be carefully utilized in the context of critical mMTC to balance the requirement trade-off between the URLLC's high access priority and massive access from mMTC. On one hand, preamble reservation (also known as semi-persistent-scheduling (SPS)) for URLLC is preferred to avoid access contention and guarantee its access priority. Nevertheless, in most cases of the coexistence of URLLC and mMTC, it may become resource-utilization inefficient and reduce the preamble availability for accommodating massive sporadic traffic from mMTC. On the other hand, although contention-based grant-free RA is more flexible and efficient in preamble resource utilization, it may cause preamble/data collisions between devices that transmit simultaneously over the shared resource, potentially jeopardizing the transmission reliability and latency. Thus, new challenges are imposed on preamble resource utilization in grant-free RA for critical mMTC, which are to manage preamble collisions across heterogeneous device types while meeting the access requirement of each individual service.

### C. Dynamic Heterogeneous Resource Optimization

In critical mMTC, it is crucial to effectively support data transmissions of both services while taking URLLC as a high priority. Due to limited radio resources, it is essential to accommodate their co-existence by jointly considering their contrasting specifications and requirements in terms of bandwidth, density, latency, and reliability. Thus, how to effectively orchestrate wireless resources in a dynamic and intelligent manner under different levels of service requirements is a challenging task.

### D. Wide Area Coverage (WAC)

Wide areas are common environments in mMTC, but ill-suited for URLLC. This poses challenges to critical mMTC

in terms of latency and reliability while making conventional URLLC solutions infeasible. Indeed, preserving massive preambles and increasing transmit power for critical mMTC devices is infeasible subject to limited bandwidth and energy, calling for novel techniques and network architecture design principles.

In addition to time and frequency diversities, spatial diversity needs to be exploited for resolving the constraints in critical mMTC services. However, accurate CSI acquisition for spatial diversity comes at the cost of feedback overhead which should be reduced to manageable level.

### E. Difficulties at Network Level

The major difficulties from networking prospective for the critical mMTC can be outlined as follows:

*a)* Need for highly flexible and scalable networking: Owing to the need to flexibly and dynamically serve a great number of dynamically connected devices ranging from a factor of ten to hundred times relative to existing cellular networks, sustaining network efficiency, QoS for URLLC service with increased mMTC network density is a non-trivial challenge.

*b)* Congestion avoidance and management: The incorporation of critical mMTC framework for the existing 5G-based cellular network will result in congestion at different levels of the network including RAN, the core network and the signaling system. As a result, the careful monitoring and management for congestion control in the converged network system is a challenging task.

*c)* Context-aware Distributed Computation and Caching: With the increasing demand of context aware and time sensitive communications in vertical domains, it is essential to investigate synergies among computing and caching resources for critical mMTC as they are distributed across different devices in such massive IoT networks. However, the standard approach to cellular network management is sluggish in terms of network resource management, and a new decentralized approach is required.

## VII. POTENTIAL CRITICAL MMTC ENABLERS

Delay-sensitive and delay-tolerant devices can co-exist as discussed in [107] [102] [108]. In this section, we discuss potential enablers for critical mMTC where URLLC (or delay-sensitive) devices and non-URLLC (or delay-tolerant) devices co-exist with different requirements.

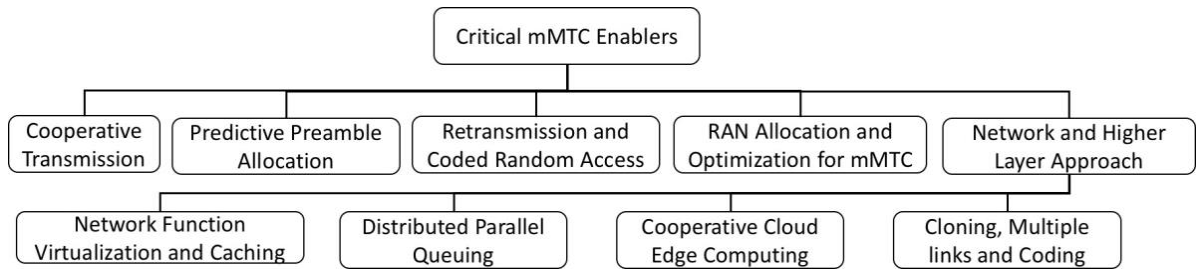

Fig. 4: Potential enablers for critical mMTC of future cellular IoT networks.

## A. Cooperative transmissions

Cooperative transmission provides spatial diversity when devices cannot be equipped with multiple antennas. See [109] and the references therein for details. While MTC devices are equipped with single antenna, base stations require high received SNR for critical and reliable communications. Especially, a delay-sensitive device can implement virtual uplink MIMO by achieving the cooperative transmission with certain delay-tolerant devices. Both remaining battery life and geometric distances may be considered in selecting cooperative devices.

## B. Predictive Preamble Allocation

When delay-sensitive and delay-tolerant devices co-exist, preamble allocation to different groups of devices is crucial to meet requirements, especially for delay-sensitive devices. To this end, as in [107] [108], optimal approaches for preamble allocation can be employed for given devices' activities and properties to meet the requirements. However, in practice, it is often difficult to know such information. Thus, new predictive algorithms can be developed.

For example, some delay-sensitive devices may have pseudo-periodic activities that allow a scheduler to predict their activity and perform predictive preamble allocation using learning-based intelligent algorithms. This can result in a low access delay without severe degradation in terms of the spectral efficiency.

## C. Re-Transmission Strategies and Coded RA

In RA with preamble transmissions, re-transmissions are often inevitable due to a finite number of preambles. To shorten access delay, a short backoff delay can be used for re-transmission strategies. In particular, the backoff delay can be zero and immediate re-transmission can also be allowed [110].

Unfortunately, immediate re-transmission results in the increase of total system load. In other words, the number of URLLC or delay-sensitive devices has to be limited to keep the system stable with guaranteed access delay. To increase the number of devices, the notion of coded RA [111] can be adopted into immediate re-transmission.

## D. RAN Allocation and Optimization

A number of research activities have discussed the co-existence of heterogeneous services in a shared physical infrastructure from the RAN resource allocation perspective [6], [112], [113]. Non-orthogonal sharing of RAN resources in uplink communications and heterogeneous NOMA can be developed as potential critical mMTC enablers [6]. By taking advantage of different reliability requirements for different services, reliability diversity can then be introduced as a design principle across the services in order to ensure performance guarantees with non-orthogonal RAN slicing. A context aware risk-sensitive formulation may be considered to allocate resources to the delay sensative devices while minimizing the risk of the mMTC transmission but ensuring URLLC delay/reliability [112]. The URLLC-mMTC co-existence can also be considered by employing power-domain NOMA within a shared resource block, where each sub-carrier can be shared by delay sensitive and delay tolerant devices [113]. Moreover, we may formulate a joint sub-carrier and transmission power allocation problem to maximize the number of successfully connected critical mMTC devices that can satisfy their QoS requirements. Another promising approach with RAN slicing for mMTC and bursty URLLC services may adaptively orchestrate the resources for critical mMTC devices [114]. In addition, a multilevel MAC scheduler can also be investigated to abstract (and intelligently share) the physical resources among network slices [115].

With relevant insights from the aforementioned works and findings discussed earlier, we have outlined the following important ideas for future RAN slicing research.

- RAN slicing under time-varying channel. There is a significant trade-off between number of resource blocks and level of modulation and coding scheme - to ensure desired QoS requirements for each (mobile/virtual) network operators. We anticipate that with higher modulation and coding scheme, higher is the spectral efficiency, and may require small number of resource blocks. Such an approach could be useful for plausible handling of the critical mMTC requirements
- Cross-slice spectrum sharing will be a promising research direction to improve the RAN slicing-based future Cellular networks. In the standard 5G network, the spectrum management and sharing is usually performed in a homogeneous environment for services of one type only either mMTC or URLLC (and within the slice). However, in the anticipated cross-slice spectrum sharing network, the spectrum sharing mechanism could be more efficient (but highly challenging) when happens across RAN slices facilitating heterogeneous service requirements of critical mMTC applications.

○ Utility optimal theoretic approach to quantify the price/loss for the network operator while designing RAN slices could offer essential QoS of the critical mMTC even with several network operators and under adverse network conditions, e.g., congested networks with multiple (mobile/virtual) network operators, networks under attacks or failures etc.

○ Adaptive slicing and resource sharing. As mentioned, existing RAN slice may consist of dedicated and shared resources, e.g., in terms of processing capabilities, memory, and so on and is fully separated from the other network slices and their resources. For critical mMTC applications, in future cellular networks, one may think redesigning an efficient approach to jointly slice and share the resources in an adaptive fashion.

### E. Network and Higher Layers Approach

We propose the following four main ideas for from a higher layer perspectives for the coexistence of delay-sensitive and delay-tolerant devices.

*a)* Network Function Virtualization and Caching for RAN: As noted, the service heterogeneity of critical mMTC domains demands diverse QoS requirements. To address the aforementioned challenges of critical mMTC, aggregate network capacity can be abstracted and sliced into multiple virtual networks by exploiting NFV. Moreover, the SDN with NFV not only will facilitate the data and control decoupling but also can provide programmable interface for a network via central controller. With a comprehensive view for the physical network, the controller often enables the efficient utilization of network resources even in time-varying network traffic and channel conditions.

Furthermore, one can divide one physical IoT network into multiple (virtual) networks based on service needs of devices, where the SDN controller can intelligently allocate and update the available system resources among these virtual networks based on their URLLC and/or mMTC requirements. Among all of these virtual networks, each mMTC device can be associated with one virtual network to access the underlying IoT network. In addition to the NFV, it is often beneficial to enhance the aforementioned ideas by using complementary techniques such as intelligent computing and caching from other networks.

*b)* Distributed Parallel Queuing: The standard approaches to enhance the 5G RACH control performance are mainly based on the ALOHA-based approach which are known to suffer from inefficiency, uncertainty and instability issues [29]. One naive approach to this end can be a distributed parallel queuing approach along the lines of that of distributed queuing collision avoidance [116], [117]. It guarantees distributed and stable performing paradigm for critical mMTC. Moreover, multiple distributed queues can operate in parallel, where first queue, can be used for collision resolution for resolving access-request signal collisions, while the other queue, can be used for data transmission and so on for other queues based on the desired set of functionalities. The intrinsic gains due to distributed parallel queuing will be: i) elimination of the back-off periods and avoidance of collisions in data packet transmissions; ii) independence in performance irrespective of the number of transmitting nodes; iii) stabilization of the traffic conditions; and iv) utilization of few bits for control signaling.

*c)* Cooperative Cloud-Edge Computing: Cloud computing platform has very high temporal and spatial capacity useful for mMTC applications. It provides a comprehensive framework for the underlying network but due to the round trip delay from nodes to the clouds, it may not directly applicable for critical mMTC applications. On the other hand, edge-computing and caching are suitable for URLLC but poses lower computing capacity and memory. To this end, a tightly coupled hybrid cloud and edge computing approach is required for a coordinated and load balanced operation, cloud helping edge (and vice versa) satisfying heterogeneous service requirements of critical mMTC. We anticipate that such a cooperative hybrid approach will address the diverse natural challenges of a mix of delay-sensitive and delay-tolerant devices .

*d)* Clustering with Data Segregation: By categorizing critical mMTC devices into several clusters based on suitable services (such as URLLC and geolocation requirements) and then segregating the device packets at the gateways, the RAN congestion can be ameliorated [118], [119]. In addition, such investigation provides novel insights into energy-efficient clustering along with segregated data sets and may facilitate the implementation of critical NB-IoT and delay-sensitive devices.

*e)* Packet Cloning, Multiple Connectivity and Network Coding: By considering packet cloning over multiple links as an example, we can always send specially coded packets over several links such that the loss of a coded packet sent over one link can be compensated by another coded packet received from a different link, and the critical mMTC device does not have to wait for a long time to detect and retransmit the lost packet as well. All coded packets will be potentially identical in terms of information content and the numbering and ordering of packets will also become unimportant. Thus, in addition to compensating for the packet losses through data redundancy, controlled cloning and coding can essentially eliminate the notion of sequencing as well. The robustness can be improved not only due to the use of multiple links but also due to the redundancy of the data sent over multiple links. In fact, network coding has been a celebrated approach for effective handling of decentralized communications and caching services. With respect to critical mMTC applications, network coding can always make it easier for the network operator to improve the efficiency of its data storage against caching, packet transmission or computing nodes failures. Data reliability can be improved substantially with replication, which may be in the form of coded information or basic cloning of packets [120].

Looking forward, conventional block codes, such as the Reed-Solomon codes, can significantly reduce storage costs compared to the storage required by using caching and cloning approach. In general, all information has to be transmitted across the network to retrieve missing information, leading to a memory-repair traffic tradeoff. To this end, network coding can help at optimal points by balancing such memory-repair traffic tradeoff. It will also improve information security and may support memory-repair trading despite other expensive approach at the cellular infrastructure.

## VIII. CONCLUDING REMARKS

It is anticipated that next generation cellular IoT networks will support massive connectivity for resource-constrained devices while meeting their diverse and critical QoS requirements. We need to address a number of challenges in order to improve performance such as scalability, reliability and latency. Within this article, we include a literature review of mMTC and URLLC to enable essential mMTC. To this end, critical mMTC is characterized by mMTC to support both delay-sensitive (or URLLC devices) and delay-tolerant devices. Since two different types of devices co-exist in a system, we have studied diverse requirements for heterogeneous features of critical mMTC in this paper. In particular, existing approaches have been reviewed and key enabling technologies from different perspectives are also identified.

## IX. ACKNOWLEDGMENTS

This research work has been supported by the ETRI, Korea, for April - November, 2020.